\documentclass{llncs}

\usepackage[usenames,dvipsnames]{color}
\usepackage{amsmath}
\usepackage{algorithmic}
\usepackage{enumerate}
\usepackage{algorithm}
\usepackage{amssymb}
\usepackage{subfigure}
\usepackage{multirow}
\usepackage{graphicx}
\usepackage{epstopdf}
\usepackage{hyperref}
\usepackage{xspace}
\usepackage{lipsum}
\usepackage{ulem}
\usepackage{bigstrut}
\usepackage{wasysym}
\usepackage[utf8]{inputenc}
\usepackage{tikz}
\usepackage{pgfplots}

\graphicspath{{./floats/eps/}{./floats/pdf/}{./floats/png/}{./floats/jpg/}}
\DeclareGraphicsExtensions{.eps,.pdf,.png,.jpg,.jpeg}

\begin{document}


\title{Analysis of friendship network from MMORPG based data}

\author{Dean Črnigoj}

\institute{University of Ljubljana, Faculty of Computer and Information Science,\\
SI-1001 Ljubljana, Slovenia, Email:
\email{dc9932@student.uni-lj.si}}

\maketitle


\begin{abstract}
This work analyzes friendship network from a Massively Multiplayer Online Role-Playing Game (MMORPG). The network is based on data from a private server that was active from 2007 until 2011. The work conducts a standard analysis of the network and then divides players according to different groups based on their activity. Work checks how friendship network can be correlated to the clan (a self-organized group of players who often form a league and play on the same side in a match) network. Main part of the work is the recommendation method for players that are not part of any clan and it is based on communities of friendship network.

\end{abstract}



\section{\label{sec:intro}Introduction}
The virtual world is for sure different than the real world, but the data may not be so different between this two worlds \cite{phillips2013researching}. The network we analyze comes from a world where every player can have one or more characters that play in this virtual world. Their friends may be friends from the real world or friends they met in this game due to same game habits, timezone, start of playing and other. Being friends in the virtual world give them some benefit, but it's not really necessary to be friends to play together.

The network data we are using is from a private server running Lineage II\footnote{Lineage II, massive multiplayer online role-playing game (MMORPG) -  \url{http://www.lineage2.com/} .} emulations software. The data was collected from a server that was online from May 2007 to June 2011. The network is undirected and two characters are friends if one added the other as friend using special command. That enables them to gain better communication no matter where in the virtual world they are. Since the data does not only include friendship status, we can compare the network to other factors like character online time, kills, level, status. Using these factors we divide characters according to their activity groups and study if the results are correlated with the structure of their friendship network. Due to the fact that many characters in a friendship network do not have a clan, we propose a method for recommending best suitable clan for a character.


\section{\label{sec:related}Related work}

In \cite{Son:2011:MSN:2043164.2018496} the authors compare how different networks (including friendship) in AION\footnote{AION Online, MMORPG based game developed by NCSoft which also developed Lineage II - \url{http://www.aiononline.com} .} game act as a social network and what are their properties. Of all networks compared, the friendship network have the highest Pearson correlation coefficient with in-degree and out-degree and reciprocity \cite{lee1988thirteen}, which was higher than 0.95. That means that their directed network should act similar as our undirected network \cite{6113094}.

Another interesting work described in \cite{Jia:2015:SGR:2835206.2736698} analyzed different multiplayer online games (MOG). Their conclusion is that networks quickly become disconnected if we remove top nodes and that friendship has a positive influence on user interactions.

In \cite{hardcorecolectoion:2009} the authors divide players of Social Games according to three groups. The groups were formed based on their interactions, similarity as in our friendship network. 

\section{\label{sec:methods}Analysis}

\subsection{Network analysis}

We collected the network from daily backup and convert it to simple edge list network. We noticed that after we removed blocking users the network becomes undirected since links are always directed in pairs. The database used to extract the network has 129998 listed characters. Out of them we have 20252 characters that claimed friendship (15.6\%).

The network itself contains 31371 links which makes mean degree $<k>$ of 3.1. This means that a small portion of all available characters claimed friendship and are part of the network. Also mean degree is small which means the network is not dense.

In basic analysis of the network, we calculated connected components, large connected component, PageRank \cite{ilprints361}, betweenes centrality \cite{Brandes01afaster} and  community structure \cite{2015PhRvX...5a1027D}. We use a power-law distribution as the comparison baseline, since node degrees in many social networks are found to follow this distribution \cite{Mislove:2007:MAO:1298306.1298311,Wilson:2009:UIS:1519065.1519089,Liu:2012:ESN:2339530.2339693}. We use the maximum likelihood estimation method \cite{Clauset:2009:PDE:1655787.1655789} to perform power-law curve fitting for node degrees in the friendship network.

\begin{align}
  \gamma &= 1 + n \Bigg [ \sum_{i=1}^{n}ln \frac{x_i}{x_{min}} \Bigg ]^{-1}
\label{eq:mle}
\end{align}

\begin{figure}[H]
\centering
\begin{tikzpicture}
\begin{axis}[ xlabel=Degrees k, ylabel=Cumulative Distribution P$_k$,width=\textwidth*0.7,legend pos=outer north east,ymode=log,domain=1:100,xmode=log,log basis y={10},log basis x={10}] 
\addplot[color=blue,mark=*,only marks]  coordinates {(1,0.47861939561524786)(2,0.195881888208572)(3,0.10063203634208967)(4,0.06152478767529133)(5,0.039699782737507405)(6,0.02666403318190796)(7,0.018220422674303773)(8,0.014171439857791823)(9,0.01180130357495556)(10,0.007801698597669366)(11,0.006221607742445191)(12,0.004888406083349793)(13,0.004987161761801304)(14,0.003900849298834683)(15,0.0031108038712225955)(16,0.0027157811574165516)(17,0.0022713806043847522)(18,0.0020738692474817303)(19,0.001431957337546909)(20,0.001431957337546909)(21,9.875567845151097E-4)(22,8.888011060635987E-4)(23,0.0010369346237408652)(24,8.394232668378432E-4)(25,5.925340707090658E-4)(26,3.456448745802884E-4)(27,4.4440055303179935E-4)(28,8.394232668378432E-4)(29,3.456448745802884E-4)(30,2.468891961287774E-4)(31,4.937783922575548E-4)(32,3.456448745802884E-4)(33,3.9502271380604387E-4)(34,1.4813351767726645E-4)(35,3.9502271380604387E-4)(36,2.962670353545329E-4)(37,9.875567845151097E-5)(38,2.468891961287774E-4)(39,3.456448745802884E-4)(40,9.875567845151097E-5)(41,9.875567845151097E-5)(42,4.9377839225755484E-5)(43,0.0)(44,1.9751135690302193E-4)(45,1.4813351767726645E-4)(46,0.0)(47,0.0)(48,1.4813351767726645E-4)(49,0.0)(50,4.9377839225755484E-5)(51,4.9377839225755484E-5)(52,0.0)(53,9.875567845151097E-5)(54,0.0)(55,4.9377839225755484E-5)(56,4.9377839225755484E-5)(57,0.0)(58,9.875567845151097E-5)(59,4.9377839225755484E-5)(60,1.4813351767726645E-4)(61,9.875567845151097E-5)(62,4.9377839225755484E-5)(63,0.0)(64,2.468891961287774E-4)(65,4.9377839225755484E-5)(66,0.0)(67,9.875567845151097E-5)(68,4.9377839225755484E-5)(69,0.0)(70,0.0)(71,0.0)(72,4.9377839225755484E-5)(73,0.0)(74,4.9377839225755484E-5)(75,4.9377839225755484E-5)(76,0.0)(77,0.0)(78,0.0)(79,4.9377839225755484E-5)(80,0.0)(81,0.0)(82,4.9377839225755484E-5)(83,0.0)(84,0.0)(85,0.0)(86,0.0)(87,0.0)(88,0.0)(89,0.0)(90,4.9377839225755484E-5)(91,0.0)(92,0.0)(93,0.0)(94,0.0)(95,0.0)(96,0.0)(97,0.0)(98,0.0)(99,0.0)(100,0.0)}; 

\addplot[color=green,mark=x,only marks]  coordinates {(1,0.14112186450720918)(2,0.2199288959115149)(3,0.2208176970175785)(4,0.1703535453288564)(5,0.10952004740272565)(6,0.05525380209362038)(7,0.025034564487458028)(8,0.009233655935216275)(9,0.003258937388899862)(10,0.0012838238198696426)(11,1.4813351767726645E-4)(12,9.875567845151097E-5)(13,0.0)(14,0.0)(15,0.0)(16,0.0)(17,0.0)(18,0.0)(19,0.0)(20,0.0)(21,0.0)(22,0.0)(23,0.0)(24,0.0)(25,0.0)(26,0.0)(27,0.0)(28,0.0)(29,0.0)(30,0.0)(31,0.0)(32,0.0)(33,0.0)(34,0.0)(35,0.0)(36,0.0)(37,0.0)(38,0.0)(39,0.0)(40,0.0)(41,0.0)(42,0.0)(43,0.0)(44,0.0)(45,0.0)(46,0.0)(47,0.0)(48,0.0)(49,0.0)(50,0.0)(51,0.0)(52,0.0)(53,0.0)(54,0.0)(55,0.0)(56,0.0)(57,0.0)(58,0.0)(59,0.0)(60,0.0)(61,0.0)(62,0.0)(63,0.0)(64,0.0)(65,0.0)(66,0.0)(67,0.0)(68,0.0)(69,0.0)(70,0.0)(71,0.0)(72,0.0)(73,0.0)(74,0.0)(75,0.0)(76,0.0)(77,0.0)(78,0.0)(79,0.0)(80,0.0)(81,0.0)(82,0.0)(83,0.0)(84,0.0)(85,0.0)(86,0.0)(87,0.0)(88,0.0)(89,0.0)(90,0.0)(91,0.0)(92,0.0)(93,0.0)(94,0.0)(95,0.0)(96,0.0)(97,0.0)(98,0.0)(99,0.0)(100,0.0)}; 

\addplot[color=red, dashed] {x^-2.22323429316};

\legend{Friendship degrees, Random degrees,$ P_k \sim  k^{-\gamma}$};
\end{axis}
 \end{tikzpicture}
\caption{Cumulative distribution of degrees in friendship and random network. }
\label{plot:friendn}
\end{figure}
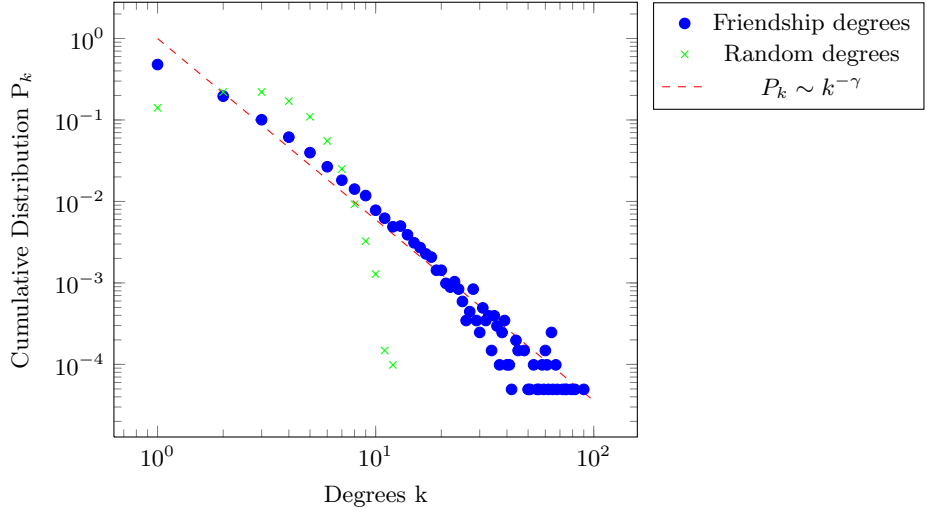

 From the nature of the network we expect that the network is scale-free and sparse.  From calculated power-law exponent we get $\gamma=$2.22, which is the reason for being scale-free network as stated in \cite{barabasi2003scale}. Figure \ref{plot:friendn} shows cumulative distribution of degrees in friendship and random network, generated with same number of nodes and links. The power-law function is dashed with red dots.

The strongest connected component contains 15339 nodes (75.7\%) and 28193 (89.9\%) links. The clustering coefficient is 0.1. The average shortest path in the strongest connected component $<d>$ is  6.3. We can say that the network is small-world \cite{Watts1998} since \( <d> \, \sim \frac{ln \, n}{ln <k>} \) and clustering coefficient of an random network based on same number of  nodes and links is 0.0002 which is much lower than original clustering coefficient.

\subsection{Character groups}

As in \cite{hardcorecolectoion:2009} we divided characters according to three different groups (\textit{Hardcore, Casual, Peripheral}). We divided them based on removing highest degree node and checking while size of strongest connected component is larger than the number of disconnected players. The basic principle is shown in Alg. \ref{alg:hardcore}.

\begin{algorithm}[H]
\caption{Calculate groups}
\begin{algorithmic}
\REQUIRE Network $n$
\ENSURE Hardcore, Casual, Peripheral  
\STATE \COMMENT{SCC represents nodes in strongest connected component.} 
\WHILE{$size(SCC(n)) \geq size(DISS(n))$}
\STATE Add(BestDegree(n),Hardcore) 
\STATE Remove(BestDegree(n),n)
\ENDWHILE
\STATE $Peripheral \leftarrow  DISS(n)$      \COMMENT{DISS represents all disconnected nodes.}  
\STATE Remove(DISS(n),n)
\STATE $Casual \leftarrow  n$  
\end{algorithmic}
\label{alg:hardcore}
\end{algorithm}

We also consider running PageRank algorithm and divide players based on their score. In both cases we compared results to characters online time and kills, and check for correlation.

We remove nodes with highest degrees until the size of the strongest connected component is larger than number of disconnected nodes as stated in Alg. \ref{alg:hardcore}. With this procedure we get 1474 \textit{Hardcore} players (7\%), 15901 \textit{Casual} players (78\%) and  2877 \textit{Peripheral} players (14\%). The results may be somehow strange since there are more \textit{Casual} players than \textit{Peripheral} players, but if we compare the numbers to characters that did not claimed friendship and mark them as \textit{Peripheral} we get \textit{Hardcore} players (1\%), \textit{Casual} players (12\%) and \textit{Peripheral} players (87\%).

The coleration between \textit{Hardcore} players and players with most online time is 0.35. We compare the same \textit{Hardcore} players with highest kills and we get 0.39. The result is first strange but then obvious since players who kill more players have more interaction online than the ones that just hunt animals and may have higher online time.
Using PageRank \cite{ilprints361} we get even better results 0.37 with online time comparison and 0.42 with  highest kills. Betweenes centrality \cite{Brandes01afaster} gives similar results as PageRank.

\subsection{Clan recommendation}

Out of 20252 characters only 6113 have clans (30.2\%). Due to high number of characters without clan the main focus of this work is a recommendation method that will connect more characters in clans. 

The usage of friendship network came in it's communities. We use Infomap community detection method\cite{PhysRevX.5.011027} between generated communities and clans players had. The result was 0.308 NMI (Normalized Mutual Information), which is relatively low and expected since this is direct comparison between clans and communities. We use players with without clans in result and we repeated measure only with players that had claimed clans. The result is 0.67 NMI which give information that we can extract clans using friendship network.

The proposed method first checks which characters do not have a clan, marks them and tries to find an appropriate clan based on their friends. The search look for players in the same community and check which clan has most players from that community. If the clan is full or does not meet any other criteria the players from clan are removed from the network and procedure repeats. The basic principle with checks for full clans is shown in Alg. \ref{alg:reccomendation}.

\begin{algorithm}[H]
\caption{Recommend clan}
\begin{algorithmic}
\REQUIRE Network $n$, Player p
\ENSURE Clan
\REPEAT
\IF{ $getClan(p) $}
\STATE{ $Clan \leftarrow getClan(p)$ }    \COMMENT{Player already have a clan.}  
\ELSE 
 \STATE $C \leftarrow ReturnCommunity(n,p)$ \COMMENT{Return players that are in same community with p.}  
  \FORALL { Players in C}
   \STATE \COMMENT{Count which clan have most players from C.}  
   \ENDFOR
    \IF{ $ size(clanWithMostPlayers) \geq $ MAX\_CLAN\_SIZE }
    \STATE RemovePlayersFromClan(clanWithMostPlayers,n)   \COMMENT{If clan is full remove players that are in a clan from a network and repeat procedure.} 
    \ELSE 
      \STATE{ $Clan \leftarrow clanWithMostPlayers$ }
    \ENDIF
 \ENDIF
\UNTIL{Clan is set}

\end{algorithmic}
\label{alg:reccomendation}
\end{algorithm}

The purposed method should first check for clans that does not reach requirements, but we think it is better that we remove clan by clan so we do not destroy network structure and we remove only those who are checked by players community. In the final proposed method the check for clans also include checking which clan has most points and does not include them in the recommended clan so that we keep balance in the gameplay.

\section{\label{sec:conclusion}Conclusion}

Considering whole players base the analyzed network represent only small portion of players in the server.  Even with that small portion we are able to extract many valuable data from it. We prove that the network is small-world and scale free and it has all properties of an social network. Based on the network we are able to divide players into different groups based on their activity. The clan recommendation method based on network communities resulted in valuable player connection between clans and friendship and it will be implemented in one of next version of server software.





\bibliographystyle{splncs03}

\end{document}